\documentclass[aps,prl]{revtex4}
\usepackage{graphicx}
\usepackage{bm}

\begin{document}
\preprint{cond-mat/0000000}

\title{Fluctuations in $YBa_{2}Cu_{3}O_{6.5}$ single crystal: an evidence for $%
2D\rightarrow 3D$ crossover}

\author{B. Rosenstein}
\altaffiliation[Permanent address: ]{Nat. Center for Theoretical
Sciences and Electronic Physics Department, Nat. Chiao Tung
University, Hsinchu, Taiwan, R.O.C.}
\author{B.~Ya.~Shapiro}
\thanks{Corresponding author}
\author{R.~Prozorov}
\altaffiliation[Present address: ]{Loomis Laboratory of Physics,
University of Illinios at Urbana-Champaign, IL 61801, U.S.A.}
\author{A.~Shaulov}
\author{Y.~Yeshurun}
\affiliation{Institute for Superconductivity, Department of Physics,\\
Bar-Ilan University, 52900 Ramat-Gan, Israel}
\date{\today }

\begin{abstract}
Magnetization measurements as a function of temperature are reported for $%
YBa_{2}Cu_{3}O_{6.5}$ crystal $\left( T_{c}=45.2\ K\right) $ for
fields between $0.2$ to $3.5\ Tesla$. All isochamps for $H>$ $1$
$Tesla$ intersect at $T_{2D}^{\star }\simeq 42.8\ K$, implying
fluctuations contribution to the magnetization. These curves
collapse into a single curve when magnetization and temperature
are scaled according to the predicted ''$2D$ scaling'' in the
fluctuation regime. Surprisingly, the low field curves also
intersect, at $T_{3D}^{\star }\simeq 43.4\ K$, and they obey a
$3D$ scaling. We provide a theoretical picture of the
magnetization in the fluctuation regime based on the
Lawrence-Doniach model. Within this model we calculate the field
and temperature dependence of the magnetization. The two
intersection points and the $2D\rightarrow 3D$ crossover are
consistent with the experimental observation.


\end{abstract}
\maketitle

\section{Introduction}

High-temperature superconductors (HTS) are characterized by a wide
temperature range in which fluctuations are important \cite{wahl97,junod}.
This range\ is proportional to the Ginzburg parameter $Gi$ , which is very
sensitive both to the dimensionality $D$\ of the system and to the
superconducting coherence length $\xi $. Thus, strong fluctuations, usually
negligible in conventional superconductors, become extremely important due
to the small $\xi $\ and quasi two-dimensional structure.

A useful tool in the analysis of the nature of critical fluctuation is the
dimensionality dependent scaling of the magnetization $M$ versus temperature
$T$ for various $DC$ magnetic fields $H$ \cite{Tesanovic92} in part of the
phase diagram that is ''not very far'' from $H_{c2}(T)$ \cite{Lawrie}. The
scaled magnetization $m=M/\left( HT\right) ^{\left( D-1\right) /D}$ is
plotted versus the scaled temperature $a_{T}=\left( T-T_{c}\left( H\right)
\right) /\left( HT\right) ^{\left( D-1\right) /D}$ and all isochamps are
predicted to collapse onto a single curve according to a dimensionality of
the system \cite{triscone,li}. Once such a scaling is found, the
''fluctuation'' dimensionality $D$ can be determined. In $2D$ systems the
relative contribution of the fluctuations to the magnetization is much
larger than in $3D$ systems. Experimentally, the scaling approach has been
widely employed to study the highly anisotropic, quasi $2D$, $Bi-Sr-Ca-Cu-O$
$\left( 2223\right) $ \cite{li} and $\left( 2212\right) $ \cite
{kogan,Bulaevskii92,triscone,kadowaki}, where two-dimensional scaling was
shown to work very well. The same $2D$ scaling seems to work also for $%
Tl-Ca-Ba-Cu-O$ $\left( 2223\right) $ \cite{thompson,wahl}, $\left(
2201\right) $ \cite{triscone,zuo}, $\left( 2212\right) $ \cite{triscone,wahl}
and $\left( 2223\right) $ \cite{wahl}, and for $Hg-Ba-Ca-Cu-O$ $\left(
1223\right) $\cite{bae}, $\left( 1201\right) $ \cite{thompson}, and $\left(
1212\right) $ \cite{huang}.

The contribution of fluctuations to the magnetization is also borne out in
the experiment as a crossing point of all isochamps at a temperature $%
T^{\ast }$ \cite{Tesanovic92,kogan,Bulaevskii92}. At this temperature $M$ is
independent of $H$ for a large range of fields. This feature was previously
observed experimentally in $2D$ systems where $2D$ scaling was expected \cite
{triscone,li,kadowaki,thompson,wahl,zuo,bae,huang}. Theoretically there is
evidence that in both 2D and 3D the intersection is not perfect, the
intersection points however are very close to each other, especially in $2D$
\cite{Sasik}. Though most HTS are either $2D$\ or $3D$\ materials, there is,
in principle, a possibility that both $3D$\ and $2D$\ behaviors would be
measured at the same sample, depending on magnetic field and temperature.
Such $2D$ to $3D$ crossover in vortex fluctuations is expected for highly
anisotropic superconductors, at high temperatures, simply because the
coherence length $\xi $ diverges as $T$ approaches the transition
temperature $T_{c}$. An evidence for a temperature induced crossover was
found in magnetization measurements in $YBa_{2}Cu_{3}O_{7}$\cite{baraduc}.
Another possible experimental approach to study the $2D\rightarrow {}3D$
crossover may be based on the expected change in the anisotropy caused by
changing the oxygen content in $Y-Ba-Cu-O$. In this system, the anisotropy
increases with the decrease in the oxygen content. Indeed, a $3D$ scaling
was observed in a fully oxygenated $YBa_{2}Cu_{3}O_{7}\ $single crystal \cite
{salem}, but a $2D$ scaling was demonstrated \cite{Gomis} in $%
YBa_{2}Cu_{3}O_{6.6}$.

In the present work we establish, experimentally and theoretically, the
existence of a $2D$ to $3D$ crossover in the nature of fluctuations in a
high $T_{c}$ superconductor. Specifically, in the oxygen deficient $%
YBa_{2}Cu_{3}O_{6.5}$ $\left( T_{c}\approx 45.15~K\right) $ single crystal,
at high fields (above $1~Tesla$), the magnetization isochamps intersect at
one temperature $T_{2D}^{\star }$ (Fig.1) and exhibit a $2D$ type of
scaling. However, at the lower fields we find another, somewhat smeared,
crossing point at $T_{3D}^{\star }$ (Fig.2), and the magnetization exhibits
a $3D$ scaling. The presence of two intersection points, as well as $2D$ and
$3D$ scaling of magnetization in different field regimes, are the new
experimental points in this work. If one defines a field dependent
''intersection point'' as an intersection of two lines for close magnetic
fields, one observes that as field is lowered, the intersection point first
''sits'' at $(M_{2D}^{\star },T_{2D}^{\star })$, than jumps quickly to $%
(M_{3D}^{\star },T_{3D}^{\star })$ and nearly stops there. Preliminary
discussions of these results were presented in Ref. \cite{Beijing}. In the
current paper we provide a\ theoretical picture of the fluctuation in
different regimes of the field-temperature ($H-T$) plane showing a $%
2D\rightarrow 3D$\ crossover in accordance with the experimental results. We
calculate the magnetization of the Lawrence - Doniach model describing
layered superconductors using the ''bubble'' diagram resummation analogous
to that established earlier in the $2D$ and $3D$ limiting cases \cite
{RuggeriThouless}. The results are compared in the $2D$ and $3D$ limits and
also between these limits where scaling does not hold.

\section{Experiment}

Details of sample preparation are given in \cite{erb}.{\em \ }The
magnetization of the $2.45\times 3.85\times 0.8\ mm^{3}$ $%
YBa_{2}Cu_{3}O_{6.5}$ single crystal was measured by a {\it ''Quantum
Design'' SQUID} magnetometer. The high temperature part of the magnetization
($46-200\ K$) was fitted to a Curie law, $M=\chi H=(\chi _{0}+C/T)H$, and
extrapolated to temperatures below $T_{c}$. The extrapolated values of $M$
were subtracted from the raw data measured below $T_{c}$. This procedure was
repeated for each value of the applied magnetic field. In Fig. $1$ we show
the temperature dependence of the magnetization for various magnetic fields $%
H>$ $1\ Tesla$. All these curves intersect at $T=T_{2D}^{\star }\simeq
42.8~K $, indicating fluctuational contribution to the magnetization \cite
{Tesanovic92,kogan,Bulaevskii92}. The subscript $2D$ is justified by the
success of the $2D$ scaling procedure described in Fig. $3$ where we plot $%
M/\left( HT\right) ^{1/2}$ {\it versus} $\left( T-T_{c}\left( H\right)
\right) /\left( HT\right) ^{1/2}$ for magnetization curves between $1$ and $%
3\ Tesla$.

Low-field measurements ($H<1\ Tesla$) are shown in Fig. $2$. Another
intersection point, at $T=T_{3D}^{\star }\simeq 43.4~K$, is found in this
field range. This\ group of curves can be scaled by using the $3D$ scaling
procedure. This is demonstrated in Fig. $4$ where we plot $M/\left(
HT\right) ^{2/3}$ {\it versus} $\left( T-T_{c}\left( H\right) \right)
/\left( HT\right) ^{2/3}$ for magnetization curves between $0.2$ and $0.75\
Tesla$. These observations imply a $2D\rightarrow 3D$ crossover in the
vortex fluctuation regime of our sample.

\section{Theory}

In order to find domains of different fluctuation behavior in the $H-T$\
phase diagram one has to calculate the fluctuational part of the
magnetization $M$\ defined by the partition function $Z:$\
\begin{equation}
M=-\frac{1}{4\pi }\frac{\partial F}{\partial H};\qquad F=-T\ln Z;\qquad
Z=\int D\psi D\psi ^{\ast }\exp (-{\cal H}_{LD}/T)  \label{z}
\end{equation}
where $F$\ is a free energy. In the general case of a layered superconductor
with Josephson inter-layer coupling the Hamiltonian $H_{LD}$\ has the well
known Lawrence-Doniach (LD) form{\it \ \cite{LD}:}
\begin{eqnarray}
{\cal H}_{LD} &=&\sum_{n}N(\epsilon _{F})\int d^{2}{\bf r}\left[ \xi
_{ab}^{2}\left| \left( -i{\bf \nabla }-\frac{2e}{c\text{%
h\hskip-.2em\llap{\protect\rule[1.1ex]{.325em}{.1ex}}\hskip.2em%
}}{\bf A}\right) \psi _{n}\right| ^{2}+\frac{\gamma }{2}\left| \psi
_{n}-\psi _{n+1}\right| ^{2}+\right.  \label{h} \\
&&\left. (t-1)\left| \psi _{n}\right| ^{2}+\frac{\beta }{2}\left| \psi
_{n}\right| ^{4}+\frac{1}{8\pi }({\bf \nabla \times A})^{2}\right] .
\nonumber
\end{eqnarray}
The magnetic field is assumed to be constant and oriented perpendicularly to
the layers $(xy)$ and its fluctuations neglected. We use the Landau gauge: $%
{\bf A=(}0,Hx,0)$ . In Eq.(\ref{h}) $\psi _{n}(x,y)$ is the order parameter
in the$\ n$-th layer, $N(\epsilon _{F})$ is (the $2D$) density of states
within the layer, $\xi _{ab}$ is the in-plane coherence length, $\beta $ -
the Ginzburg-Landau (GL) coefficient, $\gamma =(\xi _{c}/d)^{2}$ is a
dimensionless parameter describing the inter-layer coupling, where $d$ is
the inter-layer spacing and $t=T/T_{c}$. This Hamiltonian describes the
strong in-plane superconducting fluctuations and their inter-layer
interactions and can manifest both $3D$\ and $2D\;$behavior in limiting
cases as shown below.

The nonlinear term $\left| \psi _{n}\right| ^{4}$\ in the Hamiltonian
becomes very important in the temperature range $|1-t|\sim Gi$\ at the broad
region near the transition line, preventing an exact solution of the
Hamiltonian in this regime. We are able, however, to apply various
approximations, as described below, and to show that free energy $F$\ can
exhibit a $3D\rightarrow 2D$\ crossover as temperature or field are changed.
We then are able to approximate the intersection point of the magnetization
curves in both $2D$ and $3D$ limits.

The magnetic moment of fluctuations described by the Hamiltonian (\ref{h})
may be approximately found in three limiting cases: {\em (i)} The $3D$ $XY$
model {\it \cite{Mikitik}}, {\em (ii)} the $3D$ lowest Landau levels ($LLL$)
approximation, and {\em (iii)} the $2D$ $LLL$ approximation \cite
{RuggeriThouless}.{\em \ }Case (i) is not relevant for our experiments since
it applies to too low magnetic fields. Cases (ii) and (iii) are studied
here. In the region of strong fluctuations it is convenient to expand the
order parameter in terms of the Landau levels eigenfunctions \cite{Ikeda89}
\begin{eqnarray}
\psi _{n}^{N}({\bf r}) &=&\sum_{k,q}\phi _{k,q}^{N}({\bf r})a_{k,q}^{N}
\label{n1951} \\
\phi _{k,q}^{N}({\bf r}) &=&\frac{1}{\sqrt{L_{z}L_{y}}}\left( \frac{2eH}{\pi
\text{%
h\hskip-.2em\llap{\protect\rule[1.1ex]{.325em}{.1ex}}\hskip.2em%
}c2^{N}N!}\right) ^{1/4}H_{N}\left( x-\frac{q\text{%
h\hskip-.2em\llap{\protect\rule[1.1ex]{.325em}{.1ex}}\hskip.2em%
}c}{2eH}\right) \exp \left[ iqy+ikdN-\frac{eH}{\text{%
h\hskip-.2em\llap{\protect\rule[1.1ex]{.325em}{.1ex}}\hskip.2em%
}c}\left( x-\frac{q\text{%
h\hskip-.2em\llap{\protect\rule[1.1ex]{.325em}{.1ex}}\hskip.2em%
}c}{2eH}\right) ^{2}\right]
\end{eqnarray}
where $N$ stands for Landau level number and summation index{\em \ }$q${\em %
\ }bears in mind degeneration of the{\em \ }$LLL${\em \ }state.

In the field and temperature ranges of the experiment one can rely on the $%
LLL$ approximation \cite{Lawrie}. Therefore, we retain in the Hamiltonian
only terms with $N=0:$%
\begin{equation}
{\cal H}_{LD}=\sum_{q,k}\mid a_{k,q}\mid ^{2}\left[ a_{H}+\gamma (1-\cos kd)%
\right] +\frac{\beta }{2}P
\end{equation}
where
\[
a_{H}=\frac{2eH}{\text{%
h\hskip-.2em\llap{\protect\rule[1.1ex]{.325em}{.1ex}}\hskip.2em%
}c}\xi _{ab}^{2}+t-1
\]
is a dimensionless temperature parameter, and
\[
P=\sum_{k_{i}q_{i}}I(k_{1},k_{2},k_{3},k_{4})a_{k_{1},q_{1}}^{\ast
}a_{k_{2},q_{2}}^{\ast }a_{k_{3},q_{3}}a_{k_{4},q_{4}}.
\]
Here
\[
k_{1}+k_{2}=k_{3}+k_{4}
\]
\[
q_{1}+q_{2}=q_{3}+q_{4}
\]
$I(k_{1},k_{2},k_{3},k_{4})=\int d^{2}{\bf r}\prod_{i=1,2,3,4}\phi
_{k_{i,}q_{i}}({\bf r})$.

Since we are interested in vortex liquid phase we will use the renormalized
high temperature expansion proposed for the $2D$ and $3D$ cases by Thouless
and coworkers \cite{RuggeriThouless}. The first step is to perform a
summation of all the ''bubble'' diagrams. This is equivalent to a kind of
mean field approximation in which $\left| \psi _{n}\right| ^{4}->\left| \psi
_{n}\right| ^{2}\left\langle \left| \psi _{n}\right| ^{2}\right\rangle .$ In
this approximation the Hamiltonian (\ref{h}) becomes

\begin{equation}
{\cal H}_{LD}^{MF}=\sum_{q,k}\mid a_{k,q}\mid ^{2}\left[ a_{H}+\gamma
(1-\cos kd)+\frac{\beta }{2}\sum_{q,k}\frac{\left\langle \mid a_{k,q}\mid
^{2}\right\rangle }{L_{x}L_{y}}\right] \frac{N(\epsilon _{F})}{d}
\label{ham}
\end{equation}
The dimensionless average
\[
\Delta \equiv \frac{\beta }{2}\sum_{q,k}\frac{\left\langle \mid a_{k,q}\mid
^{2}\right\rangle }{L_{x}L_{y}}
\]
can be found self-consistently by solving the gap equation:

\begin{equation}
\Delta =\frac{Td\beta }{2L_{x}L_{y}N(\epsilon _{F})}\sum_{q,k}\frac{1}{%
a_{H}+\gamma (1-\cos kd)+\Delta }=\frac{2T\beta eH}{N(\epsilon _{F})\pi
\text{%
h\hskip-.2em\llap{\protect\rule[1.1ex]{.325em}{.1ex}}\hskip.2em%
}c}\frac{1}{\sqrt{(a_{H}+\gamma +\Delta )^{2}-\gamma ^{2}}}  \label{ham1}
\end{equation}

The magnetization calculated from Eq.(\ref{z}), using the mean field
Hamiltonian (\ref{h}), is
\begin{equation}
M=-\frac{eN(\epsilon _{F})\xi _{ab}^{2}}{\pi \beta \text{%
h\hskip-.2em\llap{\protect\rule[1.1ex]{.325em}{.1ex}}\hskip.2em%
}cd}\Delta .  \label{ham2}
\end{equation}
For convenience, we convert the expression for $\Delta $ into a
dimensionless form
\begin{equation}
\Delta =g\frac{bt}{\sqrt{(b+t+\Delta +\gamma -1)^{2}-\gamma ^{2}}},
\label{ham3}
\end{equation}
where
\[
b=\frac{H}{H_{c2}(0)};\text{ }g=\frac{2T_{c}\beta eH_{c2}(0)}{N(\epsilon
_{F})\pi \text{%
h\hskip-.2em\llap{\protect\rule[1.1ex]{.325em}{.1ex}}\hskip.2em%
}c}.
\]
The dimensionless coupling constant $g$ is proportional to the $2D$ Ginzburg
number.

We consider two limits for which the gap equation can be solved {\it %
analytically}. The first case refers to the domain of the $H-T$ phase
diagram where

\begin{equation}
\gamma <<\frac{1}{2}\left( \sqrt{(b+t-1)^{2}+4bt}+(b+t-1)\right)
\label{ham3a}
\end{equation}
namely, for the experiments at relatively high fields. In this case, a well
known $2D$\ result (see \cite{Lawrie}) reads:
\begin{eqnarray}
\Delta &=&(\sqrt{bt})f_{2D}(u);\text{ }u=\frac{b+t-1}{2\sqrt{bt}},\text{ }
\label{ham4} \\
&&\text{ \ }f_{2D}(u)=\sqrt{u^{2}+1}-u  \nonumber
\end{eqnarray}
where
\[
b+t-1=\frac{T-T_{c}(H)}{T_{c}}
\]
The magnetization in this limit demonstrates a well pronounced $2D$ scaling
dependence
\begin{equation}
\frac{M}{\sqrt{HT}}\propto f_{2D}\left( \frac{T-T_{c}(H)}{\sqrt{HT}}\right)
\label{ham4a}
\end{equation}

The second case refers to the limit
\begin{equation}
\gamma >>b+t-1+\frac{2((b+t-1)^{3}}{27},\text{ and }V=\frac{%
4((b+t-1)^{3}\gamma }{27g^{2}(bt)^{2}}>\frac{1}{2}  \label{ham4b}
\end{equation}
namely for the experiments at relatively low fields. In this case,
\begin{equation}
\Delta =[(bt)^{2/3}]\left( \frac{g^{2}}{\gamma }\right) ^{1/3}f_{3D}(V)
\label{ham5}
\end{equation}
where
\[
f_{3D}(V)=2V^{1/3}\sin (\frac{\pi }{6}-\frac{\varphi }{3})-\left( \frac{V}{2}%
\right) ^{1/3}
\]
and
\begin{equation}
\text{ }\tan \varphi =\frac{\sqrt{2V-1}}{1-V}.  \label{ham6}
\end{equation}
Apparently, the behavior of the magnetization in this case is caused by $3D$
fluctuations:
\begin{equation}
\frac{M}{(HT)^{2/3}}\propto f_{3D}\left( \frac{T-T_{c}(H)}{(HT)^{2/3}}\right)
\label{ham6a}
\end{equation}
In both limits one clearly finds a scaling behavior, manifested in Figures 3
and 4. For an intermediate magnetic field, however, scaling is not expected
even though the LLL approximation is still valid. In this intermediate case
the scale is provided by the inter-layer spacing $d$.

\section{Fits and discussion}

The solid lines in Figures 1, 2, and 5 are the theoretical magnetization
{\it vs.} temperature curves derived from Eq. (\ref{ham4}) for the $2D-$%
behavior (Fig. 1), from Eq. (\ref{ham6}) for the $3D$ behavior (Fig. 2), and
from Eq. (\ref{ham2}) for the intermediate case (Fig. 5). Also, as implied
by Eqs. (\ref{ham4a}) and (\ref{ham6a}), in both the $2D$ and the $3D$ cases
the magnetization data is expected to scale. Both of these formulas obey the
respective scaling conditions as demonstarted by the solid lines in Figures
3 and 4 derived from Eqs. (\ref{ham4}) and (\ref{ham5}), respectively. All
the $"2D"$ experimental curves intersect at about $T_{2D}^{\star
}~/~T_{c}\simeq 0.948$ and the $3D$ curves approximately intersect at
somewhat higher temperature $T_{3D}^{\star }~/~T_{c}\simeq 0.960$. Table I
summarizes the superconducting parameters used in our analysis to fit the
experimental data.\bigskip

\begin{table}
\caption{Parameters used for calculations of theoretical curves in
Figs. $1$ , $2$, and $5$.}
\begin{tabular*}{6cm}{|@{\extracolsep{0ptplus1fil}}l@{\extracolsep{0ptplus1fil}}|l@{\extracolsep{0ptplus1fil}}|}
\colrule
Parameter & Value used \\
\colrule

$T_{c}$ & $45.15~K$ \\

$\frac{dH_{c2}}{dT}|_{T=T_{c}}$ & $4~Tesla/K$ \\

$\xi _{ab}(0)$ & $14$ $A$ \\

$\frac{\beta }{N(\varepsilon _{F})}$ & $0.65\frac{sec^{2}}{g}$ \\

$d$ & $5$ $A$ \\

$\beta _{BCS}$ & $2.6\times 10^{27}erg^{-2}$ \\

$N(\varepsilon _{F})$ & $4\times 10^{27}\frac{1}{erg\text{ }cm^{2}}$ \\

$\xi _{c}(0)$ & $0.5$ $A$ \\

$\frac{m_{c}}{m_{ab}}=\left( \frac{\xi _{ab}}{\xi _{c}}\right)
^{2}$ & $850$
\\
\colrule
\end{tabular*}
\end{table}

The parameters were derived in the following way: The transition
temperature
$T_{c}$ was derived directly from the magnetization data. The slope $\frac{%
dH_{c2}}{dT}|_{T=T_{c}}=4$ $Tesla/K$, yielding $H_{c2}(0)=180$ $Tesla$, gave
the best fit to the experimental data for both intersection points. Here we
use the notation $H_{c2}(0)$ to denote $T_{c}\frac{dH_{c2}}{dT}|_{T=T_{c}}$
rather than unknown upper critical field at zero temperature. The latter is
unknown and sometimes is estimated as 70\% of this quantity as in BCS theory
inapplicable to the present case. The coherence length was defined by $\xi
_{ab}(0)=\sqrt{\Phi _{0}/2\pi H_{c2}(0)}$. \ It should be noted that here $%
\frac{dH_{c2}}{dT}|_{T=T_{c}}$ is the mean field theoretical parameter
rather than experimentally measured; direct measurement of this value is
expected to yild a smaller value due to contribution of fluctuations \cite
{Shapiro}. The value of the dimensionless coupling constant $g=0.07$, the
{\em interlayer-coupling} parameter $\gamma =0.008$, and the magnetization $%
M_{2D}^{\ast }=-2.5\times 10^{-4}$ $emu$ at the $2D$ intersection point,
determine the ratio $\frac{\beta }{N(\varepsilon _{F})}$, $d$, and $\xi
_{c}(0).$ Assuming the validity of the BCS expression for $\beta =\frac{%
7\zeta (3)}{8\pi ^{2}T_{c}^{2}}$ one gets a rough estimate of the density of
states. Now we discuss the applicability of the $2D$ and $3D$ limits to
discribe the regions around the crossing points as was done on Fig.1 and 2.
The inequality \ref{ham3a} for $t=T^{\ast }/T_{c}=.95$, so that $1-t>>b=.001$
simplify into $b>>\gamma ^{\ast }(1-t)=.2Tesla/H_{c2}(0)$. Similarly the
condition of applicability of the $3D$ limit ( see Eq. (\ref{ham4b})) can be
simplified into $b<<1-t-\gamma =5Tesla/H_{c2}(0)$. The use of the 2D limit
in Fig.1 is therefore justified for $B=1.5Tesla$ or larger, while for $B\leq
.75Tesla$ the use of the $3D$ limit in Fig.2 is justified.

To conclude, a crossover between $2D$ and $3D$ behavior in the magnetization
of $YBa_{2}Cu_{3}O_{6.5}$ was observed and described theoretically by
employing the the Lawrence - Doniach model in lowest Landau level
approximation in the fluctuation regime. The model yields analytical
expressions for {\it two} intersection points of the magnetization curves,
as observed experimentally. One intersection point, for magnetization curves
at relatively high fields, is a result of fluctuations in the $2D$ regime.
The second intersection point, for relatively low fields, is a result of
fluctuations in the $3D$ regime. The model also predict scaling of the the
magnetization data in the $2D$ and the $3D$ regmes, as observed
experimentally.

\begin{acknowledgments} This work was supported by The Israel Science Foundation -
Center of Excellence Program, and by the Heinrich Hertz Minerva
Center for High Temperature Superconductivity. Y.Y. acknowledges
support from the U.S.-Israel Binational Science Foundation. We
thank A. Erb and H. Wuhl for providing the YBCO sample. B. S. and
A.S. acknowledge support from the Israel Academy of Science and
Humanities. B.R. is very grateful to the Physics Department at Bar
Ilan University for the warm hospitality.
\end{acknowledgments}

\begin{figure} [tb]
\includegraphics[width=8.5cm,keepaspectratio=true]{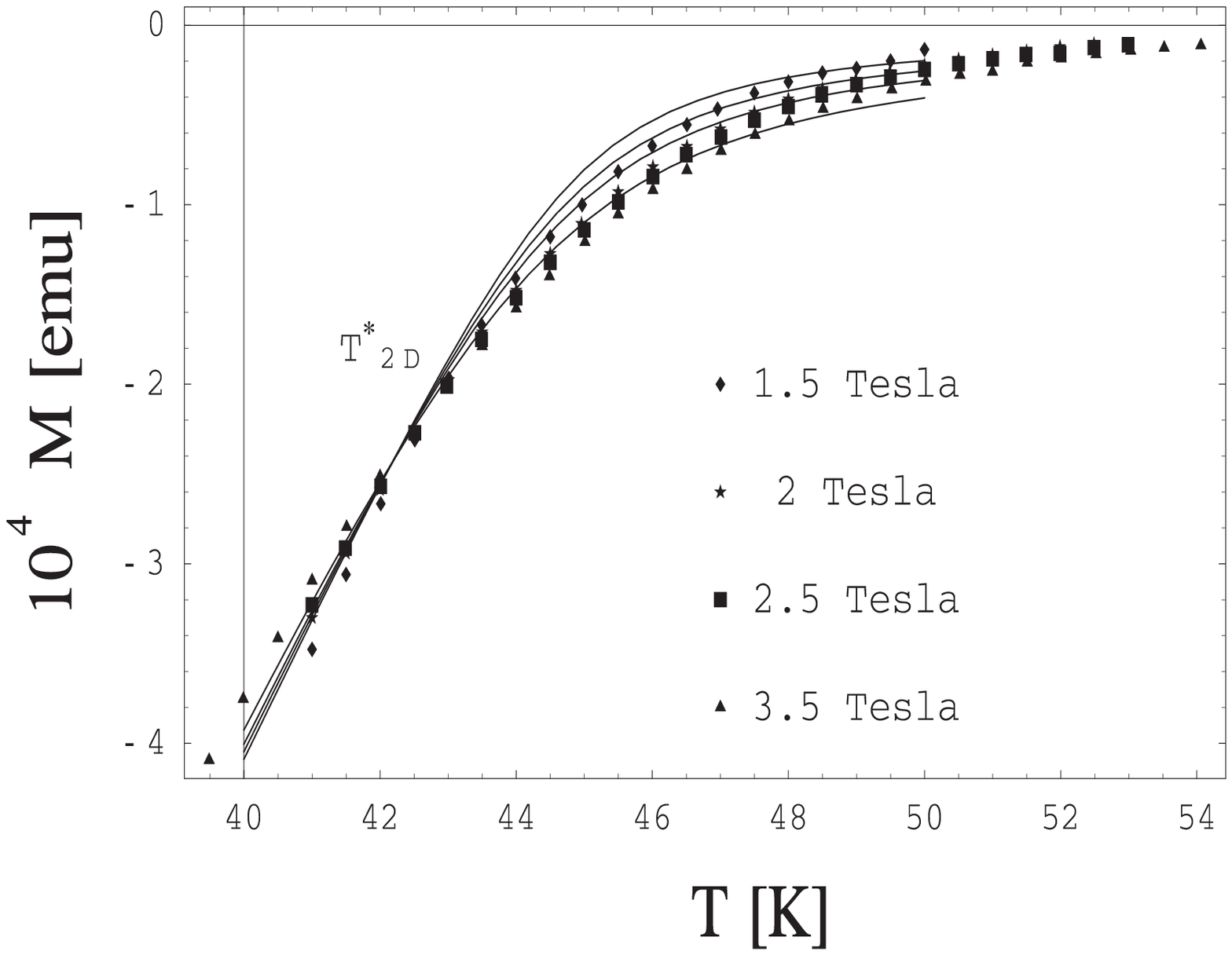}
 \caption{Magnetic moment versus temperature in the
high-field region $(1.5-3\ Tesla)$. The solid lines are fits to
Eqs. (\ref {ham2}) and (\ref{ham4}) obeying the $2D$ scaling with
the parameters listed in Table I.}
 \label{fig1}
\end{figure}
\begin{figure} [tb]
\includegraphics[width=8.5cm,keepaspectratio=true]{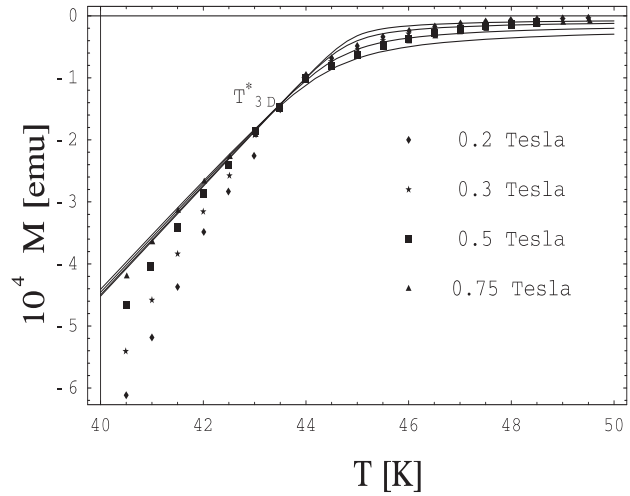}
 \caption{Magnetic moment versus temperatures in the
low-field region $(0.2-0.75\ Tesla)$. The solid lines are fits to
Eqs. (\ref {ham2}) and (\ref{ham6}) obeying the $3D$ scaling with
the parameters listed in Table I.}
 \label{fig2}
\end{figure}
\begin{figure} [tb]
\includegraphics[width=8.5cm,keepaspectratio=true]{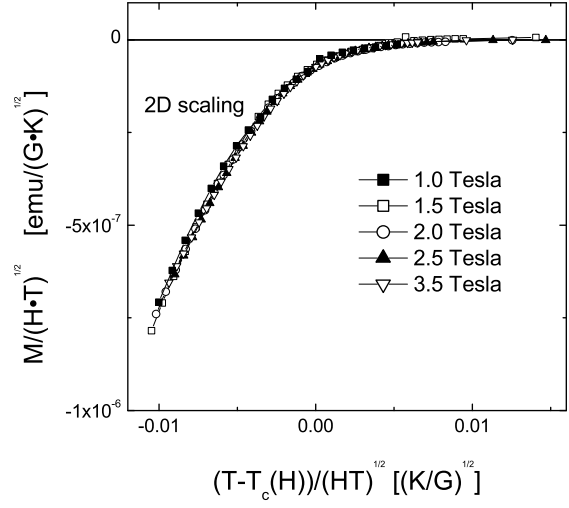}
 \caption{Two-dimensional scaling of the high-field data. The solid line is
a fit to Eq. ((\ref{ham4a})).}
 \label{fig3}
\end{figure}
\begin{figure} [tb]
\includegraphics[width=8.5cm,keepaspectratio=true]{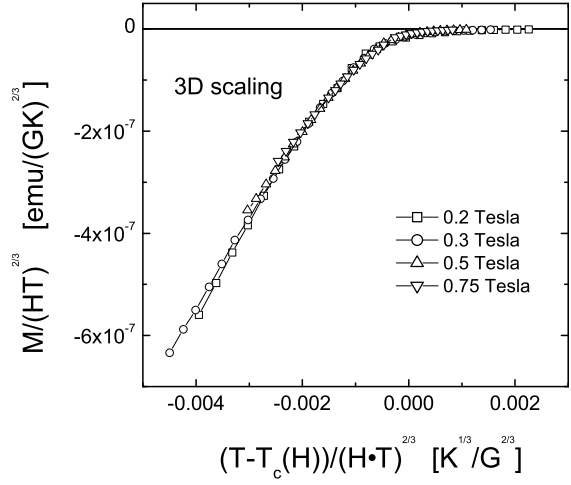}
 \caption{Three-dimensional scaling of the
low-field data. The solid line is a fit to Eq. (\ref{ham6a}).}
 \label{fig4}
\end{figure}
\begin{figure} [tb]
\includegraphics[width=8.5cm,keepaspectratio=true]{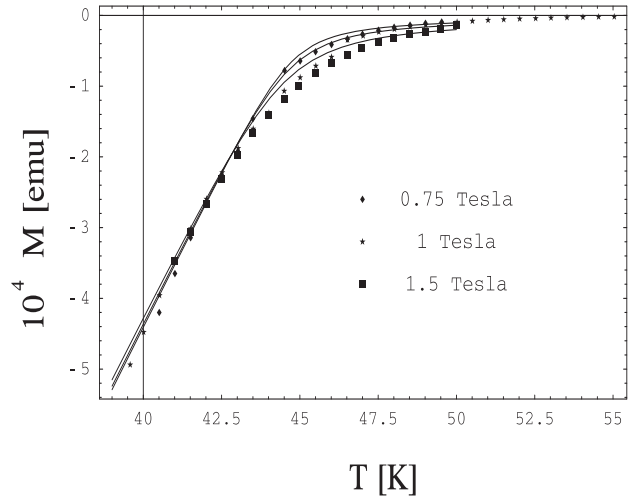}
 \caption{Magnetic moment versus temperature for various
magnetic fields around $1$ $Tesla$. This field range represents an
intermediate regime where $M(T)$ does not obey neither $3D$ nor
$2D$ scaling; It is still described, however, in the framework of
the Lawrence - Doniah model, Eqs. (\ref{ham2}) and (\ref{ham3}),
as described by the solid lines.}
 \label{fig5}
\end{figure}

\end{document}